\documentclass{JHEP3}
\usepackage{graphicx,amsmath,amssymb}
\usepackage{epsfig,multicol}
\usepackage{epsf}
\usepackage{epstopdf}
\input{epsf.sty}
\usepackage{amsmath,amssymb,rotating,wasysym,multirow,lscape,chngpage}
\usepackage{graphicx}
\def\half{{1\over 2}}
\def\({\left(}
\def\){\right)}
\def\[{\left[}
\def\]{\right]}

\def\e{\begin{equation}}
\def\q{\end{equation}}
\def\m{\begin{eqnarray}}
\def\n{\end{eqnarray}}

\title{Constraints on the primordial gravitational waves with variable sound speed from current CMB data}
\author{Cheng Cheng$^{1}$ \footnote{chcheng@itp.ac.cn}, Qing-Guo Huang$^{1}$ \footnote{huangqg@itp.ac.cn},
Xiao-Dong Li$^{1,2}$ \footnote{renzhe@mail.ustc.edu.cn}, Yin-Zhe
Ma$^{3,4}$ \footnote{mayinzhe@phas.ubc.ca}
\\\small{$^1$ \em
State Key Laboratory of Theoretical Physics, Institute of Theoretical Physics, Chinese Academy of Sciences, Beijing 100190, China}
\\\small{$^2$ \em
Department of Modern Physics, University of Science and Technology of China, Hefei 230026, China}
\\\small{$^3$ \em Department of Physics and Astronomy, University of British Columbia, Vancouver, V6T 1Z1, BC Canada}
\\\small{$^4$ \em Canadian Institute for Theoretical Astrophysics, 60 St. George Street
Toronto, M5S 3H8, Ontario, Canada}
 }

\abstract{ We make a comprehensive investigation of the
observational effect of the inflation consistency relation. We
focus on the general single-field inflation model with the
consistency relation $r=-8c_s n_t$, and investigate the
observational constraints of sound speed $c_s$ by using the Seven-Year
{\it WMAP} data, the BICEP tensor power spectrum data,
and the constraints on $f_{\rm NL}^{\rm equil.}$ and $f_{\rm
NL}^{\rm orth.}$ from the Five-Year {\it WMAP} observations. We
find that the constraints on the tensor-to-scalar ratio $r$ is
much tighter if $c_s$ is small, since a large tilt $n_t$ is
strongly constrained by the observations. We obtain $r<0.37,\
0.27$ and $0.09$ ($dn_s/d\ln k=0$) for $c_s$=$1$, $0.1$ and $0.01$ models
at $95.4\%$ confidence level. When taking smaller values of $c_s$,
the positive correlation between $r$ and $n_s$ also leads to
slightly tighter constraint on the upper bound of $n_s$ , while
the running of scalar spectral index $dn_s/d\ln k$ is generally
unaffected. For the sound speed $c_s$, it is not well constrained if only
the CMB power spectrum data is used, while the constraints are
obtainable by taking $f_{\rm NL}^{\rm equil.}$ and $f_{\rm
NL}^{\rm orth.}$ priors into account. With the constraining data
of $f_{\rm NL}^{\rm equil.}$ and $f_{\rm NL}^{\rm orth.}$, we find that, $c_s\lesssim
0.01$ region is excluded at 99.7\% CL, and the $c_s=1$ case (the single-field
slow-roll inflation) is slightly disfavored at $68.3\%$ CL. In
addition, the inclusion of $f_{\rm NL}^{\rm equil.}$ and $f_{\rm
NL}^{\rm orth.}$ into the analysis can improve the constraints on
$r$ and $n_s$. We further discuss the implications of our
constraints on the test of inflation models.}


\begin{document}

\section{Introduction}

An important task of modern cosmology is to understand the
expansion history of the Universe. The standard hot big-bang model
is successful in explaining various observations, including
Hubble expansion, Big-bang Nucleosynthesis and microwave background radiation \cite{Weinberg},
yet still suffers from the flatness, horizon and monopole
problems, etc. The inflation model, in which the vacuum energy
drives the Universe exponentially expanding in the very early
Universe \cite{inflation}, was proposed under such concerns.
Besides the successful explanation of the above problem,
inflationary cosmology can provide a viable mechanism for the
origin of the cosmic structures.

There have been numerous inflation models proposed in the last
several decades. In the face of so many competing candidates, it
is necessary to find an effective way to figure out which one is
realistic, or at least, which one is most favored by the
cosmological observations. Especially, it is important to confirm
or rule out the canonical single-field slow-roll (SFSR) inflation
model.

It has been proved that the SFSR inflation can generate observable
primordial scalar and tensor perturbations, which
encode themselves in the cosmic microwave background (CMB)
anisotropies. Thus, it is possible to test inflationary models
from the current CMB observations, e.g., the Wilkinson Microwave
Anisotropy Probe ({\it WMAP}) satellite \cite{WMAP}, QUaD
experiment \cite{QUaD}, BICEP experiment \cite{BICEP} and other
probes \cite{OtherCMB}.

There have been a number of investigations made on testing
the inflation models from the current and future CMB observations
\cite{testinflation}, mainly on the issues of constraining the SFSR
model with the scalar spectral index $n_s$, running of the spectral index
$dn_s/d\ln k$, and tensor-to-scalar ratio $r$ as free parameters. Besides the determination of the  parameters in inflation models, it has also been proposed \cite{QGH} that
the consistency relations, which features various types of inflation models, can be used as a test to classify and distinguish
different models of inflation. The possibility of the observational
test of the consistency relations has been discussed in
\cite{QGH} in detail.

In this paper we make further investigations on the observational
effect of the consistency relation $r=-8c_s n_t$. We focus on the
general single-field inflation model, and discuss the current constraints on the sound
speed $c_s$ from the CMB data, including the Seven-Year {\it WMAP}
(WMAP7) power spectrum data \cite{WMAP7}, the BICEP data
\cite{BICEP}, and the constraints on $f_{\rm NL}^{\rm equil.}$ and
$f_{\rm NL}^{\rm orth.}$ from the Five-Year {\it WMAP} (WMAP5)
observations \cite{fNL,fNL2}. We then discuss the results of the
constraints on $n_s$, $r$, $dn_s/d\ln k$ parameters when $c_s\neq
1$.

This paper is organized as follows. In Sec. 2, we introduce the
inflationary consistency relation in the general single-filed
inflation model. In Sec. 3, we briefly introduce the CMB data and
data analysis methodology used in this paper. The results of constraints on
cosmological parameters are presented in Sec. 4 and Sec. 5. We
summarize our results in Sec. 6.

\section{Single-field Inflation Model}

Let's start with the general single-field inflation model
\begin{eqnarray}
S=\int d^4 x \sqrt{-g} \[{M_p^2\over 2}R+P(X,\phi)\],
\end{eqnarray}
where $M_p=1/\sqrt{8\pi G}$ is the reduced Plack mass, $R$ is the Ricci scalar,
$g$ is the determinant of the metric,
and $X=-\half g^{\mu\nu}\partial_\mu \phi \partial_\nu \phi$.
$P(X,\phi)$ is an arbitrary function of $X$ and $\phi$.
This action is the most general Lorentz invariant action
for inflaton $\phi$ minimally coupled to Einstein gravity. The
primordial scalar power spectrum of curvature perturbation is
\cite{Garriga:1999vw}
\begin{eqnarray}
\Delta_R^2={H^2/M_p^2\over 8\pi^2c_s \epsilon},
\end{eqnarray}
where
\begin{eqnarray}
\epsilon=-{\dot H\over H^2},
\end{eqnarray}
is the slow-roll parameter, and
\begin{eqnarray}
c_s={P_{,X}\over P_{,X}+2XP_{,XX}}
\end{eqnarray}
is the speed of sound. The spectral index of scalar curvature perturbation power spectra becomes
\begin{eqnarray}
n_s-1={d\ln \Delta_R^2\over d\ln k}=-2\epsilon-\eta-s,
\end{eqnarray}
where
\begin{eqnarray}
\eta={\dot \epsilon\over H\epsilon},\ s={\dot c_s\over H c_s},
\end{eqnarray}
are another two slow-roll parameters. Due to the dynamics of inflation,
the spectral index $n_s$ can be scale-dependent as well.
Its scale-dependence is measured by the running of the spectral index $dn_s/d\ln k$.
The primordial power spectrum of scalar curvature perturbation then takes the form,
\begin{equation}\label{eq:scalarpower}
 \Delta_R^2(k)
 =\Delta^2_R(k_0)\left(\frac{k}{k_0}\right)^{n_s(k_0)-1+\frac{1}{2}dn_s/d\ln k},
\end{equation}
where $k_0$ is the pivot scale.
The primordial power
spectrum of gravitational waves perturbation generated during
inflation only depends on the Hubble parameter during inflation
\begin{eqnarray}
\Delta_T^2={H^2/M_p^2\over \pi^2/2},
\end{eqnarray}
with tilt
\begin{eqnarray}\label{eq:nt}
n_t={d\ln \Delta^2_T\over d\ln k}=-2\epsilon.
\label{nt}
\end{eqnarray}
The tensor-to-scalar ratio is defined as
\begin{eqnarray}
r=\Delta^2_T/\Delta_R^2=16c_s\epsilon,
\end{eqnarray}
and then by combing with Eq. (\ref{eq:nt}), we obtain the consistency relation
\begin{eqnarray}
r=-8c_s n_t.
\label{cons}
\end{eqnarray}
Here we should note that since the acceleration of scale factor takes the form
$\ddot a = H^2 a (1-\epsilon)$, inflation only happens if $\epsilon<1$,
therefore from Eqs.~(\ref{nt}) and (\ref{cons}), we know the valid ranges of values for $n_t$
and $r$ as
\begin{eqnarray} \label{eq:nT_r_cs}
-2<n_t\leq 0,\ \hbox{and}\ r<16c_s.
\end{eqnarray}

If $c_s=1$, the general single-field inflation model reduces to
the single-field slow-roll inflation.
But if $c_s \ll 1$, the non-trivial sound speed of inflation can
generate non-Gaussian modes of perturbation, which results in
a large non-local form of bispectrum \cite{Chen:2006nt}.
Although the non-local form of bispectrum has not been well classified,
the two most general types, equilateral type with shape size $f_{\rm NL}^{\rm equil.}$,
and orthogonal type measured by $f_{\rm NL}^{\rm orth.}$,
have been widely discussed in literatures \cite{WMAP7,fNL2,Huang:2010up}.
In \cite{Huang:2010up}, the observational
constraint on the $c_s$ from the bispectrum has been discussed
and the requirement from the stability of the field theory
($c_s^2\geq 0$) implies $f_{\rm NL}^{\rm orth.}\leq -0.054 f_{\rm
NL}^{\rm equil.}$.

\section{Data Analysis Methodology}

In the following data analysis, we will combine WMAP7 power spectrum \cite{WMAP7},
with BICEP tensor power spectrum data \cite{BICEP} and bispectrum constraints
on $f_{\rm NL}^{\rm equil.}$ and $f_{\rm NL}^{\rm orth.}$ \cite{fNL,fNL2},
to constrain inflation consistency relation.
The {\it WMAP} $TT$ power spectrum at $2\leq l \leq 1200$ is powerful to
constrain the cosmological parameters, e.g. $n_s$ and $dn_s/d\ln
k$. We also use the {\it WMAP} $TE$/$EE$ data at $2\leq l \leq
800$, and the $BB$ data mainly on large scales $2\leq l \leq 23$.
To be consistent with the {\it WMAP} results \cite{WMAP7}, we
choose our pivot scale to be $k_0=0.002\ {\rm Mpc}^{-1}$.

We also use the BICEP tensor power spectrum data which mainly
covers the region $21 \leq l \leq 335$.
%It was reported that the BICEP data can constrain $r<0.72$ (95.4\% CL) for the SFSR model \cite{BICEP}.
Following the pipelines of \cite{BICEP}, we
construct the expected bandpowers for the inflation models, and
use the lognormal approximation to calculate the $\chi^2$
function,
\begin{equation}
 \chi^2({\bf p}) = \left[\hat{{\bf Z}}^{BB}-{\bf Z}({\bf p})^{BB}\right]^T
 \left[{\bf D}^{BB}({\bf p})\right]^{-1}
 \left[\hat{{\bf Z}}^{BB}-{\bf Z}({\bf p})^{BB}\right],
\end{equation}
where ${\bf p}$ is the model parameters,
and $\hat{{\bf Z}}^{BB}$ and ${\bf Z}({\bf p})^{BB}$ are the observational and theoretical bandpowers.
${\bf D}^{BB}({\bf p})$ is the covariance matrix which is dependent on the model parameters.
The likelihood function then takes the form
\m
\mathcal{L}\propto \frac{1}{\sqrt{\det [{\bf D}^{BB}({\bf p})]}} e^{-\chi^2({\bf p})/2}.
\n

In addition, the observational results of $f_{\rm NL}^{\rm
equil.}$ and $f_{\rm NL}^{\rm orth.}$ can also constrain the value
of $c_s$. We use $f_{\rm NL}^{\rm equil.}$ and $f_{\rm
NL}^{\rm orth.}$ priors obtained from the WMAP5 observations
\cite{fNL,fNL2}, and construct the $\chi^2$ function as
\begin{equation}
 \chi^2({\bf p})=v({\bf p})_{\rm WMAP}^T C^{-1} v({\bf p})_{\rm WMAP},
\end{equation}
where $C$ is the covariance matrix given in \cite{fNL2}, and $v({\bf p})_{\rm WMAP}$
is the difference between the observed and model values of $f_{\rm
NL}^{\rm \rm equil.}$ and $f_{\rm \rm NL}^{\rm \rm orth.}$ \cite{fNL2},
\begin{eqnarray}
 v({\bf p})_{\rm WMAP} = \left(
 \begin{array}{c}
\langle\hat{f}_{\rm NL}^{\rm \rm equil.}({\bf p})\rangle - (\hat{f}_{\rm NL}^{\rm \rm equil.})_{\rm WMAP} \\
\langle\hat{f}_{\rm \rm NL}^{\rm \rm orth.}({\bf p})\rangle - (\hat{f}_{\rm \rm NL}^{\rm \rm orth.})_{\rm WMAP}\\
 \end{array}
 \right).
\end{eqnarray}
The WMAP5 data
yields to \cite{fNL2}
\footnote{The covariance matrix $C$ is dependent on the data \cite{fNL2}.
Since the WMAP7 covariance matrix $C$ has not yet been published,
we will adopt the WMAP5 covariance matrix $C$ in the following discussion.}
\begin{equation}
 f_{\rm NL}^{\rm \rm equil.} = 155\pm 140, \ \ f_{\rm NL}^{\rm \rm orth.}=-149 \pm
 110,
\end{equation}
where the errors given are the $1\sigma$ confidence level.
%We numerically marginalize the nuisance parameter $\tilde{c}_3$ in the $\chi^2$ function.

We will determine the best-fit parameters and the 68.3\%
and 95.4\% confidence level (CL) ranges by using the Monte Carlo
Markov chain (MCMC) technique. The whole set of our free
parameters is
\begin{equation}
 {\bf P}=\{\Omega_b h^2, \Omega_c h^2, \theta, \tau, n_s, dn_s/d\ln k, r, c_s, A_s, A_{SZ}\}
 \footnote{$\theta$ is the ratio of the sound horizon to the angular diameter distance;
$\tau$ is the the reionization optical depth;
$A_s$ is the primordial superhorizon power in the curvature perturbation on the pivot scale $k_0=0.002\ {\rm Mpc}^{-1}$;
$A_{SZ}$ is an SZ template normalization.}.
\end{equation}
We modify the publicly available CAMB \cite{CAMB} and COSMOMC
packages \cite{COSMOMC} to include models with $c_s$ as a free parameter,
and generate $O(10^5)$ samples for each set
of results presented in this paper.

\section{Cosmological Constraints of Fixed $c_s$ Models}

In the following sections we will discuss the cosmological interpretations of the consistency relation.
In this section we focus on three models with $c_s$ fixed as 1, 0.1 and 0.01.
The free $c_s$ model will be discussed in the next section.

\subsection{Effects of $c_s$ on the Power Spectrum}

\begin{figure}
\centering
\includegraphics[width=15cm]{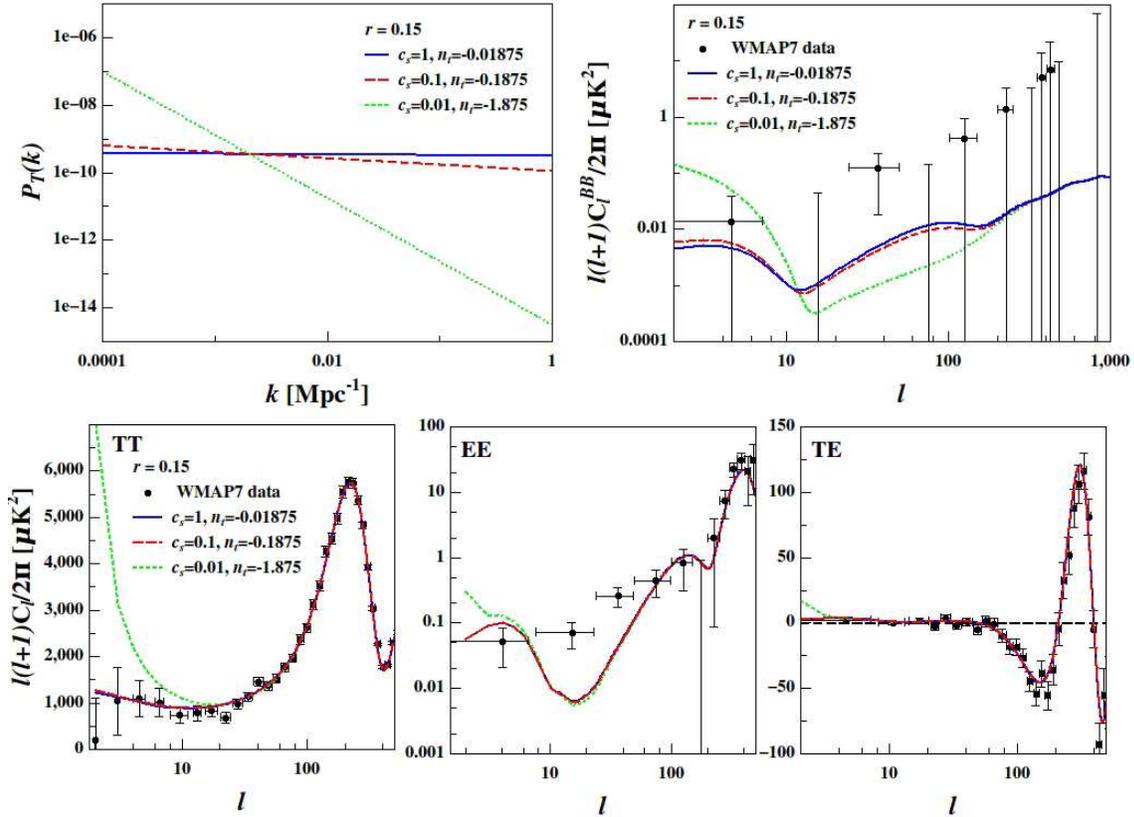}
\caption{\label{fig:powerspectra} Power spectra for inflation
models with different $c_s$. Models with $c_s$=1, 0.1 and 0.01 are
plotted in blue solid, red dashed and green dotted lines, respectively.
In all figures we fix $r=0.15$. The primordial tensor power spectrum
and the $BB$, $TT$, $EE$, $TE$ power spectra are plotted.
The {\it WMAP} data \cite{WMAP7} are plotted in black points.
The $c_s$=0.1 model with $n_t=-0.1875$ leads to slightly
larger values of power spectrum at the large scale (not very evident),
while the $c_s$=0.01 model leads to significantly larger $P_T(k)$/$C_l$s
in small-$k$/low-$l$ region.
}
\end{figure}

We firstly clarify how the different values of $c_s$ affect the shape of the angular power spectra of CMB.

In the upper-left panel of Fig.~\ref{fig:powerspectra} we plot the
primordial tensor power spectrum for $c_s$=1, 0.1 and 0.01
respectively. We see that $c_s$ has significant influence on the
tilt of the power spectrum through the consistency relation
$n_t=-r/(8c_s)$. In particular, at the large scale (small $k$), the $c_s$=0.01 and $c_s$=0.1 models have much larger
values of $P_t(k)$ than that in the model with $c_s=1$.

This effect is also visible in $BB$, $TT$, $EE$ and $TE$ power spectra,
which is shown in the upper-right and lower panels
of Fig.~\ref{fig:powerspectra} (with lensing). In all
figures we take $r=0.15$ and fix other parameters at their WMAP7
best-fit values. It is shown that the amplitude of the power
spectrum for $c_s$=0.1 is slightly larger than the $c_s$=1 case
(not evident), while for the $c_s$=0.01 case the low-$l$ $C_l$s
are significantly larger. The panels indicate that the
set of parameters $r=0.15$, $c_s=0.01$ is inconsistent with the
{\it WMAP} data. Thus, we expect a tight constraint on $r$ when
$c_s$ is small.

\subsection{Results of Fitting}

Our results of fitting for different models with fixed $c_s$ are shown in Table \ref{tab:A}.

\begin{table}[!htp]
\centering
\renewcommand{\arraystretch}{1.2}
\scriptsize{
\caption{\label{tab:A}Results of fitting with fixed $c_s$.}

\

\begin{tabular}{clccccc}
\hline\hline
\multicolumn{2}{c}{Data \& Model}   & $r$(95.4\% CL) & $n_t$(95.4\% CL) & $n_s$ & $d n_s/d\ln k$ \\
\hline \multirow{3}{*}{{\it WMAP}}
 & $c_s=1$   & $<0.37$ & $>-0.05$ & $0.967^{+0.026}_{-0.010}$ & -- \\
%\cline{2-6}
 & $c_s=0.1$ & $<0.26$ & $>-0.33$ & $0.972^{+0.016}_{-0.014}$ & -- \\
%\cline{2-6}
 & $c_s=0.01$ & $<0.09$ & $>-1.16$ & $0.966^{+0.017}_{-0.010}$ & -- \\
\hline \multirow{3}{*}{{\it WMAP}+BICEP}
 & $c_s=1$  & $<0.32$ & $>-0.04$ & $0.966^{+0.024}_{-0.008}$ & -- \\
%\cline{2-6}
& $c_s=0.1$ & $<0.26$ & $>-0.33$ & $0.971^{+0.019}_{-0.014}$ & -- \\
%\cline{2-6}
 & $c_s=0.01$ &$<0.09$ & $>-1.16$ & $0.969^{+0.014}_{-0.013}$ & -- \\
\hline \multirow{3}{*}{{\it WMAP}+BICEP}
 & $c_s=1$ & $<0.36$ & $>-0.05$ & $1.011^{+0.050}_{-0.035}$ & $-0.023^{+0.018}_{-0.024}$ \\
%\cline{2-6}
 & $c_s=0.1$ & $<0.35$ & $>-0.44$ & $1.014^{+0.070}_{-0.046}$ & $-0.022^{+0.021}_{-0.034}$ \\
%\cline{2-6}
(+$dn_s/d\ln k$)
& $c_s=0.01$ & $<0.10$ & $>-1.21$ & $1.005^{+0.068}_{-0.013}$ & $-0.019^{+0.005}_{-0.030}$ \\
\hline
\end{tabular}
}
\end{table}

Here we list the results for $c_s$=1, 0.1 and 0.01. In the 4-9
rows, the constraints on $r$, $n_t$, $n_s$ and $d n_s/d\ln k$ by
using the {\it WMAP}+BICEP data are listed, divided into the
$dn_s/d\ln k=0$ and $dn_s/d\ln k\neq0$ cases. For comparison, in
the 1-3 rows we also list the results obtained by using the {\it
WMAP} data alone with $dn_s/d\ln k=0$.
We discuss the results of constraints in Sec. 4.2.1 and 4.2.2 in detail.

\subsubsection{The $dn_s/d\ln k=0$ case}\label{secnorun}

In this subsection we briefly discuss the fitting results of models without including $dn_s/d\ln k$ as a free parameter.

\begin{figure}
\centering
\includegraphics[width=16cm]{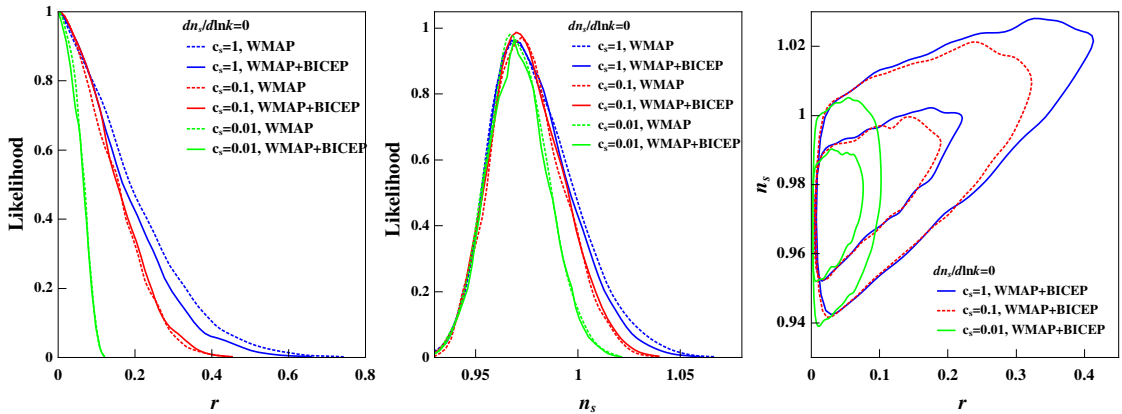}
\caption{\label{fig:fixedcs_no_nrun} Fitting results for the fixed
$c_s$ models with $dn_s/d\ln k=0$.
In the left and middle panels
we plot the likelihoods of $r$ and $n_s$ for the $c_s=1$,
0.1, 0.01 models in blue, red, green lines, and
the results obtained by using the {\it WMAP} and {\it WMAP}+BICEP are
shown in dotted and solid lines, respectively.
In the right panel we also plot the marginalized 68.3\%
and 95.4\% CL contours in the $r-n_s$ plane.
We see that smaller values of $c_s$ lead to tighter constraints on $r$,
and thus tighter upper bound constraints on $n_s$,
due to their correlation. The inclusion of BICEP data slightly improves
the constraints on $r$ and $n_s$ for the $c_s=1$ case.}
\end{figure}

Let us first see the constraints on the tensor-to-scalar ratio $r$ which determines
the amplitude of the tensor power spectrum. The fitting results of
$r$ are listed in the second column of Table \ref{tab:A} and the
likelihood functions are plotted in the left panel of Fig.~\ref{fig:fixedcs_no_nrun}. Using the {\it WMAP} data alone, we
find a 95.4\% CL constraint $r<0.37$, which is well consistent
with the result obtained by the {\it WMAP} 7-yr data ($r<0.36$). As
expected, we find the constraint on $r$ becomes much tighter
if the value of $c_s$ becomes smaller (see the left panel of
Fig.~\ref{fig:fixedcs_no_nrun}). Using the {\it WMAP}+BICEP data,
we find $r < 0.32,\ 0.26$ and $0.09$ (95.4\% CL) for $c_s=1,\ 0.1$
and $0.01$, respectively. This result is quite reasonable.
According to the consistency relation $n_t=-r/(8c_s)$, if $c_s$ is
small, a large $r$ would lead to a large $n_t$,
leading to a large tensor mode on superhorizon scale, which is
strongly constrained by low-$l$ CMB data. The necessary condition for
inflation Eq. (\ref{eq:nT_r_cs}) is automatically satisfied by the
constraint. The $r=0.15$, $c_s=0.01$ case shown in the
Fig.~\ref{fig:powerspectra} is excluded.

Except for $r$, another interesting issue is the fitting results
of the scalar spectral index $n_s$. Using the {\it WMAP}+BICEP
data, we find $n_s=0.966^{+0.024}_{-0.008}$,
$0.971^{+0.019}_{-0.014}$ and $0.969^{+0.014}_{-0.013}$ (68.3\% CL) for
$c_s=1,\ 0.1$ and $0.01$. All the results are consistent with
Harrison-Zeldovich spectrum ($n_s=1$) at 68.3\% CL. The likelihoods of $n_s$ in
different cases are plotted in the middle panel of Fig.~\ref{fig:fixedcs_no_nrun}. Similar to $r$, we find that the upper
bound constraint on $n_s$ also becomes tighter when $c_s$ is
smaller. This effect is caused by the positive correlation between
$r$ and $n_s$. In the right panel of
Fig.~\ref{fig:fixedcs_no_nrun}, we plot the marginalized
contours in the $r-n_s$ plane, which shows that the smaller $c_s$ is,
the tighter constraints on $n_s$ and $r$.

The results of constraints from BICEP data are as follows. For the $c_s$=1
model, we get $r<0.37(0.32)$ (95.4\% CL) by using the {\it
WMAP}({\it WMAP}+BICEP) data. The inclusion of BICEP data slightly
improves the constraint by $\sim 14\%$. However, for the $c_s=0.1$
and $0.01$ cases, since the constraints on $r$ mainly come from
constraints on $n_t$ by the low-$l$ {\it WMAP} data, the inclusion
of BICEP data does not lead to significant improvement in the result
\footnote{The Two-Year BICEP data, which maps only $\sim 2\%$ of the
sky, measures limited modes of perturbations, therefore does not contribute
too much on the total constraining.}. The BICEP
data can only affect $n_s$ through its correlation between $r$,
so it does not have significant effect on the results of $n_s$.

\subsubsection{The $dn_s/d\ln k\neq0$ case}

We now discuss the results of constraints with running of spectral
index $dn_s/d\ln k$ as a free parameter.

\begin{figure}[!htp]
\centering
\includegraphics[width=16cm]{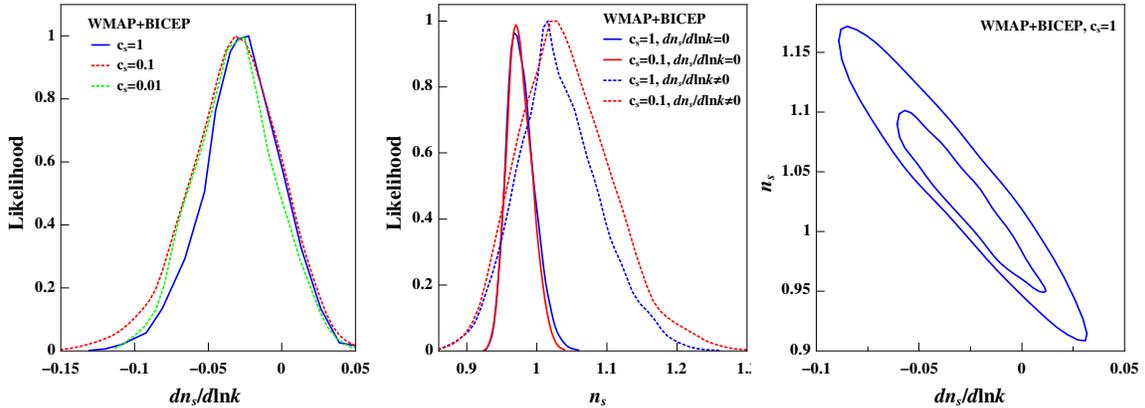}
\caption{\label{fig:fixedcs_with_nrun} Fitting results for the
fixed $c_s$ models with $dn_s/d\ln k\neq 0$.
%The $c_s=1$, 0.1 and
%0.01 cases are plotted in blue, red and green lines, respectively.
%The results obtained by using the {\it WMAP} and {\it WMAP}+BICEP
%are shown in dashed and solid lines.
The marginalized likelihoods of $dn_s/d\ln k$ and $n_s$ are shown in the left and
middle panels. In the right panel we plot the marginalized 68.3\%
and 95.4\% CL contours in the $dn_s/d\ln k-n_s$ plane. The
influence of $c_s$ on $dn_s/d\ln k$ is not very significant (left panel). The inclusion
of $dn_s/d\ln k$ significantly amplifies the allowed range of
$n_s$ and shifts its central value from below 1 to above 1 (middle panel),
since they are strongly anti-correlated (right panel; see also Eq. (2.7)).
%Be careful about the label of equation!!!!!!!!!!!!
}
\end{figure}

The fitting results are shown in the last three rows of Table
\ref{tab:A} and Fig.~\ref{fig:fixedcs_with_nrun}. For the three
models we find similar constraints on $dn_s/d\ln k$, thus $c_s$
does not have too much influence on $dn_s/d\ln k$.

The most striking effect of the inclusion of $dn_s/d\ln k$ is the
significant amplification of the allowed region of $n_s$. In the
middle panel of Fig.~\ref{fig:fixedcs_with_nrun}, we plot the
likelihood functions of $n_s$ for $dn_s/d\ln k=0$ (solid) and
$dn_s/d\ln k\neq 0$ (dotted). We see that the width of $n_s$ likelihood
is increased by nearly a factor of two. The inclusion of $dn_s/d\ln
k$ also changes the central values of $n_s$ from below 1 to above
1. These phenomena are caused by the strong anti-correlation
between $dn_s/d\ln k$ and $n_s$ (see the right panel of
Fig.~\ref{fig:fixedcs_with_nrun}).

The inclusion of $dn_s/d\ln k$ slightly releases the upper bound constraints of $r$.
For the $c_s=1,\ 0.1$, $0.01$ models,
the 95.4\% CL upper bounds on $r$ are 0.32, 0.26, 0.09 for $dn_s/d\ln k=0$
and 0.36, 0.35, 0.10 for $dn_s/d\ln k\neq0$.
%Since $dn_s/d\ln k$ also affects the tilt of the power spectrum,
%This result is reasonable.

\section{Cosmological Constraints of Free $c_s$ Models}

In this section we consider the more general case, i.e., treating
$c_s$ as a free parameter.
\footnote{We sample $c_s$ in the range of (0.001,1).}
We not only use the {\it WMAP} and BICEP power spectrum data,
but also take $f_{\rm NL}^{\rm equil.}$
and $f_{\rm NL}^{\rm orth.}$ into consideration in order to
constrain $c_s$. A summary of the fitting results, including
$c_s$, $r$, $n_t$, $n_s$ and $dn_s/d\ln k$ are given in Table
\ref{tab:B}. Notice that the word `UCON' represents for `unconstrained'.
%In the next, we will briefly introduce these results, divided into two cases, i.e., without and with $f_{\rm NL}$ including into the fit.

\begin{figure}[!htp]
\centering
\includegraphics[width=14cm]{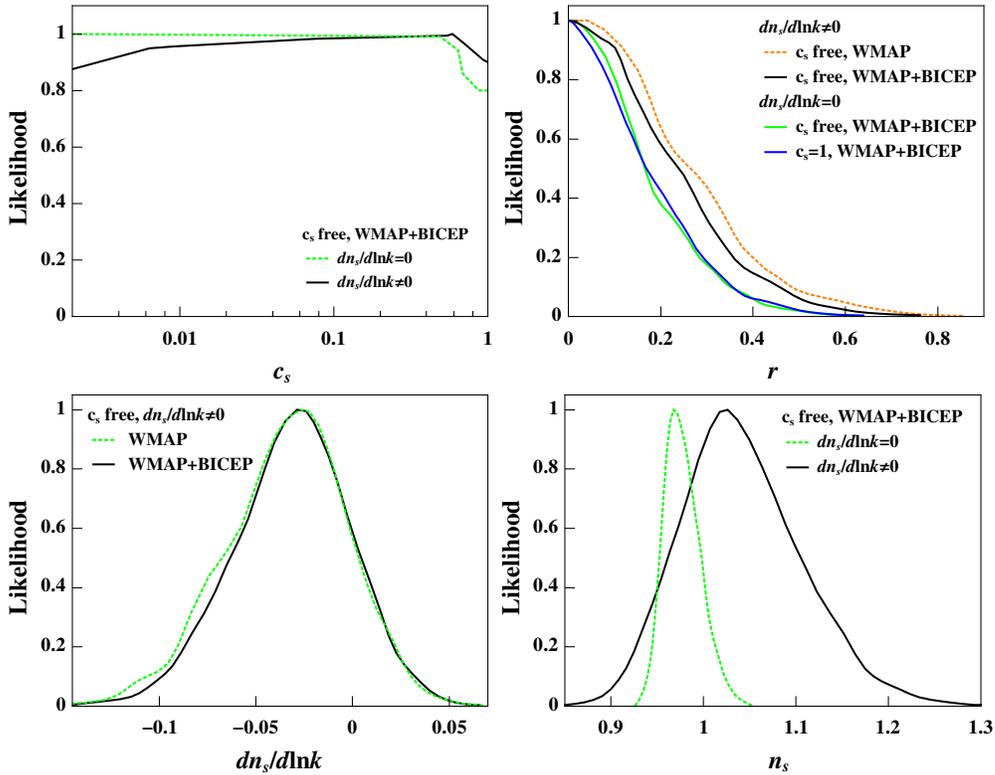}
\caption{\label{fig:freecs_like} Marginalized likelihoods
of $c_s$, $r$, $dn_s/d\ln k$ and $n_s$ for the free
$c_s$ case, obtained by using the {\it WMAP} and {\it WMAP}+BICEP data.
%The $dn_s/d\ln k=0$ and $dn_s/d\ln k\neq
%0$ cases are shown in magnetic and orange colors, respectively. As
%a comparison, the $c_s$=1 model is also plotted in the blue line.
Upper-left: The current CMB power spectrum data alone can not
constrain $c_s$. Upper-right: The inclusion of $dn_s/d\ln k$
slightly widen the distribution of $r$. Lower-left: The inclusion
of BICEP data does not affect the distribution of $dn_s/d\ln k$
too much. Lower-right: The inclusion of $dn_s/d\ln k$ greatly
boardens the width of $n_s$ likelihood, and shifts its central
value to be greater than unity.}
\end{figure}

\begin{figure}[!htp]
\centering
\includegraphics[width=15cm]{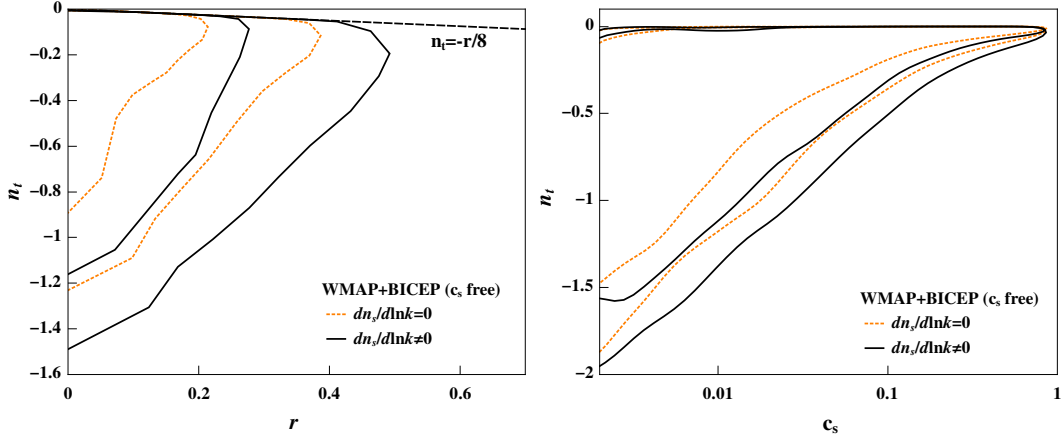}
\caption{\label{fig:freecs_rcon} Marginalized 68.3\% and 95.4\% CL
contours in the $r-n_t$ (left) and $c_s-n_t$ (right) planes which
are obtained by using the {\it WMAP}+BICEP data.
%The $dn_s/d\ln
%k=0$ and $dn_s/d\ln k\neq 0$ cases are plotted in magnetic and
%orange colors.
In the left panel, the $n_t=-\frac{r}{8}$ line is
plotted in the black dashed line.
We see that $n_t>-2$ automatically satisfied,
and the inclusion of $dn_s/d\ln k$
evidently amplifies the parameter space.}
\end{figure}

\begin{table}[!htp]
\centering
\renewcommand{\arraystretch}{1.2}
\scriptsize{
\caption{Results of fitting with $c_s$ as a free parameter.}
\label{tab:B}
\

\begin{tabular}{clcccccc}
\hline\hline
\multicolumn{2}{c}{Data \& Model}  & $c_s$ & $r$(95.4\% CL) & $n_t$(95.4\% CL) & $n_s$ & $dn_s/d\ln k$ \\
\hline \multirow{2}{*}{{\it WMAP}}
 &  & UCON & $<0.36$ & $>-1.33$ & $0.966^{+0.024}_{-0.011}$ & -- \\
 & +$d n_s/d\ln k$ & UCON & $<0.46$ & $>-1.44$ & $1.027^{+0.065}_{-0.051}$ & $-0.026^{+0.023}_{-0.030}$ \\
\hline \multirow{2}{*}{{\it WMAP}+BICEP}
 &   & UCON & $<0.32$ & $>-1.30$ & $0.967^{+0.025}_{-0.012}$ & -- \\
%\cline{2-6}
 & +$d n_s/d\ln k$ & UCON  & $<0.41$ & $>-1.40$ & $1.019^{+0.063}_{-0.038}$ & $-0.026^{+0.020}_{-0.027}$ \\
%\cline{2-6}
\hline \multirow{2}{*}{{\it WMAP}+BICEP+$f_{\rm NL}$}
 &  & $0.019^{+0.012}_{-0.006}$ & $<0.21$ & $>-0.91$ & $0.973^{+0.011}_{-0.016}$ & -- \\
 & +$d n_s/d\ln k$ & $0.016^{+0.017}_{-0.003}$ & $<0.29$ & $>-1.15$& $1.016^{+0.064}_{-0.045}$ & $-0.024^{+0.022}_{-0.033}$ \\
\hline
\end{tabular}
}
\end{table}

\subsection{Results of fitting Without $f_{\rm NL}$ Prior}

In this subsection we discuss the results obtained by {\it
WMAP}+BICEP data, without adding $f_{\rm NL}^{\rm equil.}$ and
$f_{\rm NL}^{\rm orth.}$ priors into the analysis. The fitting
results of parameters are shown in the 1-4 rows of Table
\ref{tab:B}, and the likelihoods of $c_s$, $r$, $n_s$
and $dn_s/d\ln k$ are plotted in Fig.~\ref{fig:freecs_like}.

Let us first have a look at the constraint on $c_s$, and its
likelihood is plotted in the upper-left panel of
Fig.~\ref{fig:freecs_like}. The current CMB power spectrum data is
not able to constrain on $c_s$, and the likelihood function shows
that it can take any possible value given the current constraints.

Secondly, the likelihoods of $r$ are shown in the
upper-right panel of Fig.~\ref{fig:freecs_like}. We see the
green and blue lines are close to each other, which means that the
constraint on $r$ with free $c_s$ is similar to the result for
$c_s$=1. \footnote{A similar conclusion was obtained in
\cite{Guo:2010jr}. } We get $r<0.32$ (95.4\% CL) for both $c_s=1$ and $c_s$
free models ({\it WMAP}+BICEP, $dn_s/d\ln k=0$).
%\footnote{This is reasonable, since the smaller values of $c_s$ only tighten the constraints of $r$ and $n_s$, the marginalized distribution of $r$ in $c_s=1$ and $c_s$ free cases shall look similar.}
The inclusion of BICEP data slightly improves constraint of $r$
from 0.36 to 0.32 ($dn_s/d\ln k=0$). We see the inclusion of
$dn_s/d\ln k$ as a free parameter boardens the width of distribution of $r$.
At $95.4\%$ CL, the constraint is widened from $r<0.32$ to $r<0.41$ ({\it
WMAP}+BICEP).

Thirdly, in the lower panels of Fig.~\ref{fig:freecs_like} we plot
the marginalized likelihoods of $dn_s/d\ln k$ (left)
and $n_s$ (right). We find that the results are similar to the
fixed $c_s$ models. It implies that the BICEP data almost does not
affect the constraint of $dn_s/d\ln k$, and the inclusion of
$dn_s/d\ln k$ greatly amplifies the distribution of $n_s$ and
shifts its central value to above 1.

In addition, in Fig.~\ref{fig:freecs_rcon} we plot the marginalized
2D-contours in the $r-n_t$ (left) and $c_s-n_t$ (right) planes. The
slow-roll model with  $n_t=-r/8$ is plotted in the black dashed
line in the left panel, and the $dn_s/d\ln k=0$ and $dn_s/d\ln
k\neq 0$ cases are shown in orange dotted and black solid lines,
respectively. By releasing $c_s$ as a free parameter, $n_t$ ranges to
much smaller values ($\sim$-1 to -1.5 at 95.4\% CL), especially in
the small $c_s$ region. We find $n_t>-2$ is still satisfied by the
constraints. In both panels we see the inclusion of $dn_s/d\ln k$
boardens the ranges of the parameter space.

% To see the effect of $c_s$ on the $n_s$, $r$ and $dn_s/d\ln k$,
% in Fig. \ref{fig:freecs_cscon} we plot the marginalized contours of  $c_s-n_s$ (left panel), $c_s-r$ (middle panel) and $c_s-dn_s/d\ln k$ (right panel).
% For large values of $c_s$, we see that the dependence of $n_s$/$r$/$dn_s/d\ln k$ on $c_s$ is not evident,
% while small values of $c_s$ lead to tighter constraints on $n_s$ and $r$.
% This is consistent with the results of Fig. \ref{fig:fixedcs_like},
% where we see that the for models with $c_s \ll 1$ the upper bound constraints on $r$ and $n_s$ are smaller.
% As a result, the marginalized constraints on $r$ and $n_s$ in the $c_s$-free model is similar to the constraints in the $c_s$=1 model,
% which is shown in the left and middle panels of Fig. \ref{fig:freecs_cscon}.

\begin{figure}[!htp]
\centering
\includegraphics[width=14cm]{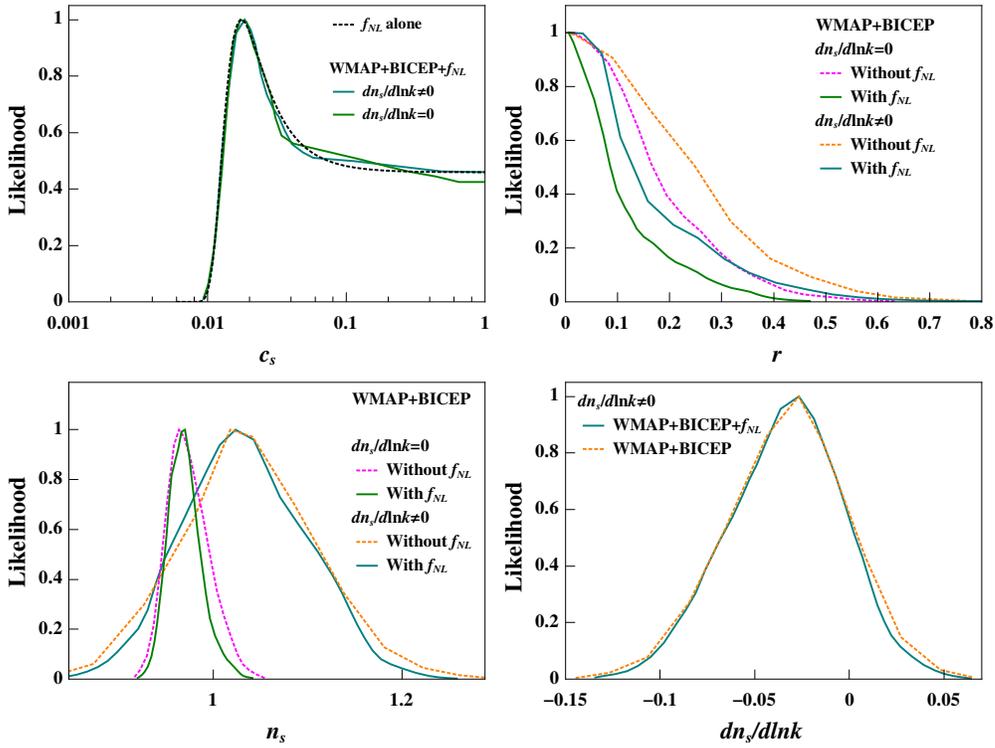}
\caption{\label{fig:fNL_like} Marginalized likelihoods
of $c_s$, $r$, $n_s$ and $dn_s/d\ln k$ for the $c_s$
free case from {\it WMAP}+BICEP data.
%The effect of $f_{\rm NL}$ priors on the constraints is clearly revealed.
%For the $dn_s/d\ln k= 0$
%case we plot the models with and without $f_{\rm NL}$ prior in
%olive and magnetic colors, while for $dn_s/d\ln k\neq 0$ the
%models with and without $f_{\rm NL}$ prior are plotted in dark
%cyan and orange colors.
%In the upper-left panel, we plot the curve
%obtained by only using $f_{\rm NL}$ prior in the black dashed
%line.
The inclusion of $f_{\rm NL}$ prior evidently tightens the
constraints on $c_s$ and $r$, and slightly improves the upper-bound
constraint on $n_s$.
The running of the spectral index $dn_s/d\ln k$ remains unchanged.}
\end{figure}

\begin{figure}[!htp]
\centering
\includegraphics[width=15cm]{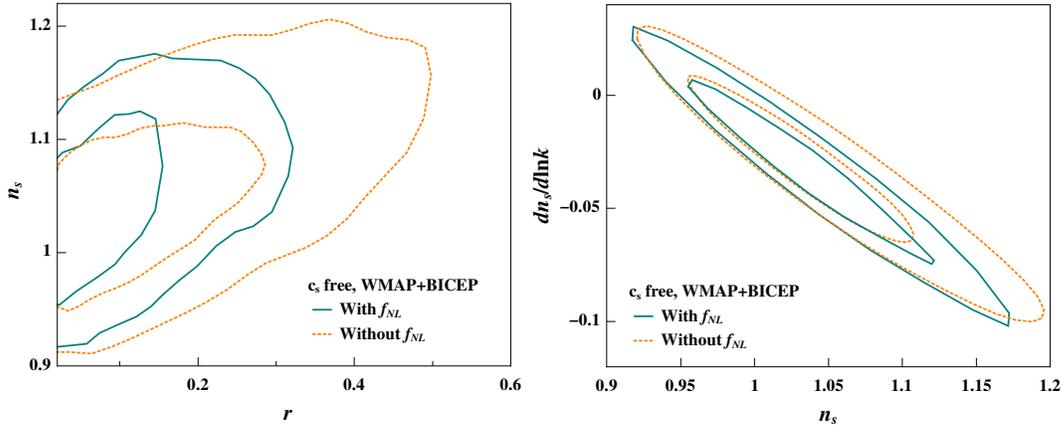}
\caption{\label{fig:fNL_con} Marginalized 68.3\% and 95.4\% CL
contours in the $r-n_s$ (left) and $n_s-dn_s/d\ln k$ (right)
planes, obtained by using the {\it WMAP}+BICEP data. In all
figures we let $dn_s/d\ln k$ and $c_s$ as free parameters. The
constraints with and without $f_{\rm NL}$ cases are shown in
dark cyan solid and orange dotted colors respectively.}
\end{figure}

\subsection{Results of fitting With $f_{\rm NL}$ Prior}

In this subsection let us take the $f_{\rm NL}^{\rm equil.}$ and
$f_{\rm NL}^{\rm orth.}$ priors into account. The fitting results
are shown in the last two rows of Table \ref{tab:B}, and the likelihoods
of $c_s$, $r$, $n_s$ and $dn_s/d\ln k$ are plotted
in Fig.~\ref{fig:fNL_like}.

The most striking effect is that the inclusion of $f_{\rm NL}^{\rm
equil.}$ and $f_{\rm NL}^{\rm orth.}$ significantly improves the
constraint on $c_s$. At 68.3\% CL, we obtain $0.013<c_s<0.031$ and
$0.013 < c_s < 0.033$ for the cases of without and with $dn_s/d\ln
k$ as a free parameter. We find the large $c_s$ region, including the SFSR inflation
with $c_s=1$, is slightly disfavored at around 68.3\% CL, while
$c_s\lesssim0.01$ is excluded at 99.7\% CL. Thus, the bispectrum data is
much more powerful than the power spectrum data for constraining
$c_s$.

By narrowing the allowed range of $c_s$, the addition of $f_{\rm
NL}^{\rm equil.}$ and $f_{\rm NL}^{\rm orth.}$ also has
interesting effect on constraining the other parameters. The
likelihood of $r$ is shown in the upper-right panel
of Fig.~\ref{fig:fNL_like}. We see that, once $f_{\rm NL}^{\rm
equil.}$ and $f_{\rm NL}^{\rm orth.}$ priors are considered, the
constraint becomes much tighter. This can be also seen through
the contours in the left panel of Fig.~\ref{fig:fNL_con}.
Again, due to the correlation between $r$ and $n_s$, where $f_{\rm
NL}^{\rm equil.}$ and $f_{\rm NL}^{\rm orth.}$ priors are included
a slightly tighter constraint on the upper bound of $n_s$ is also
obtained (see the lower-left panel of Fig.~\ref{fig:fNL_like}).
But the effect of $f_{\rm NL}^{\rm equil.}$ and $f_{\rm NL}^{\rm
orth.}$ priors on $dn_s/d\ln k$ is negligible (see the lower-right
panels of Fig.~\ref{fig:fNL_like}).

Finally, the correlations between $r$, $n_s$ and $dn_s/d\ln k$ are shown
in Fig.~\ref{fig:fNL_con}. The left panel shows the $r-n_s$
contours and the right panel shows the $n_s-dn_s/d\ln k$ contours.
In all figures we set $c_s$, $dn_s/d\ln k$ as free parameters and
use both {\it WMAP} and BICEP data. The cases of without and with
$f_{\rm NL}$ as a free parameter are shown in orange and dark cyan colors.
One can see the strong correlation between $r$ and $n_s$, which suggests
that if the distribution of $r$ is tightened, $n_s$ distribution is also constrained.
However, the distribution of running spectral index $dn_s/d\ln k$,
is not much affected by this correlation, because
the change of $n_s$ is much smaller comparing with $r$.

\section{Conclusion}

In this paper we make a detailed investigation of the cosmological
interpretation of the consistency relation from CMB data. We focus
on the general single-field inflation model in which the spectral
index $n_t$ of tensor perturbation power spectrum is related to
the tensor-to-scalar ratio $r$ by $n_t = -r/(8c_s)$, and further
investigate the effect of the sound speed $c_s$. The datasets used
in this paper include the {\it WMAP} power spectrum data, the
BICEP $B$-mode polarization data, and $f_{\rm NL}^{\rm equil.}$
and $f_{\rm NL}^{\rm orth.}$ priors obtained from the WMAP5
bispectrum data.

We discuss three models with fixed $c_s$=1, 0.1 and 0.01. We find
that when $c_s$ is small, the tilt of the tensor power spectrum
$n_t$ becomes very large if $r$ is not too small, and then
a tight constraint on $r$ is obtained for $c_s\ll 1$. Using the
{\it WMAP}+BICEP data, we obtain the 95.4\% CL constraints of
$r<$0.37, 0.26, 0.09 for the $c_s$=1, 0.1, 0.01 cases
($dn_s/d\ln k=0$). Due to the positive correlation between $r$ and
$n_s$, smaller values of $c_s$ lead to slightly tighter constraint
on the upper bound of $n_s$. The effect of $c_s$ on the running of
scalar spectral index $dn_s/d\ln k$ is not obvious. The inclusion
of $dn_s/d \ln k$ significantly alters the constraints of $n_s$,
and slightly amplifies the upper bound constraint of $r$.

For more general cases in which $c_s$ is taken as a free parameter,
we find that $c_s$ unconstrained if we only use the current CMB power spectrum data in the analysis,
and the marginalized distribution of $r$, $n_s$ and $dn_s /d \ln k$ are all similar to the $c_s=1$ case.
However, after taking $f_{\rm NL}^{\rm equil.}$ and $f_{\rm NL}^{\rm orth.}$ priors into consideration,
we find the sound speed $c_s$ is effectively constrained.
The $c_s\lesssim 0.01$ region is ruled out, and the $c_s\gtrsim0.03$ region is disfavored at the 68.3\% CL.
From the constraints on $c_s$, the inclusion of $f_{\rm NL}$ leads to tighter constraint on the $r$ and $n_s$.
In the $dn_s/d\ln k=0$ case,
we find $r<0.21/0.32$ (95.4\% CL) with/without $f_{\rm NL}$ prior ($dn_s/d\ln k=0$),
and the results for the $dn_s/d\ln k\neq0$ case is $r<0.29/0.41$.
The running of spectral index $dn_s/d\ln k$ is almost unaffected.

To summarize, we find that the consistency relation has
significant effect in the constraints on cosmological parameters
$r$ and $n_s$ when $c_s$ is small, while the parameter $dn_s/d\ln
k$ remains unaffected. Although the sound speed $c_s$ is
unconstrained by the CMB power spectrum data, it can be
effectively constrained by the CMB bispectrum data. Using the
$f_{\rm NL}^{\rm equil.}$ and $f_{\rm NL}^{\rm orth.}$ priors
obtained from the WMAP5 data, we find the SFSR model with $c_s$=1
is slightly disfavored at the 68.3\% CL. Thus,
we are expecting that, the on-going and upcoming CMB observations,
such as {\it Planck} \cite{Planck} and {\it CMBPol} \cite{CMBPol},
with much lower instrumental noise and better foreground clean,
will provide stronger constraints on inflation models.

\vspace{1.4cm}

\noindent {\bf Acknowledgments}

\vspace{.5cm}

QGH is supported by the project of Knowledge Innovation Program of Chinese Academy of Science and a grant from NSFC (grant NO. 10975167).
The MCMC is mainly performed by the Lenovo Shenteng 7000 supercomputer in the Supercomputing Center of Chinese Academy of Sciences (SCCAS).
XDL acknowledge the Institute of Theoretical Physics and the Supercomputing Center of USTC for the use of computing resources.

\vspace{1.5cm}

%\appendix

%\section{appendix}
%\label{ap}

\end{document}